\documentclass[12pt]{article}
\usepackage{graphicx}
\begin{document}
version 27.9.2013
\begin{center}
{\Large On the structure of the new particle at 126 GeV } \\ 
{ \large (Higgs- or not Higgs-boson?) }   
\vspace{0.3cm}

H.P. Morsch\footnote{permanent address: HOFF, Brockm\"ullerstr.~11,  
D-52428 J\"ulich, Germany\\ E-mail:
h.p.morsch@gmx.de }\\
National Centre for Nuclear Research, Pl-00681 Warsaw,
Poland
\end{center}

\begin{abstract}
A new particle - discovered recently with the Atlas and CMS detectors at
LHC - has been interpreted as the long sought Higgs-boson. A corresponding
scalar field is needed to make the weak interaction gauge invariant and to
understand the quark masses in the Standard Model. 

However, the Standard Model is an effective theory with quark masses, which
can be understood only in a fundamental theory. Such a theory has been
constructed, based on a generalised second order extension of QED, in which
the quarks can be understood as effective fermions with masses given by
binding energies in a boson-exchange potential. In the present approach
the Higgs-mechanism is not needed. 

In this framework a good understanding of particles in the ``top'' regime is
obtained. Two $J^\pi=1^-$ $q\bar q$ states are predicted, identified
with Z(91.2 GeV) and the $t\bar t$ state at about 350 GeV. Further, two $0^+$
$q\bar q$ states are obtained, one with a mass consistent with that of the new
particle, the other with a mass of about 41 GeV. A detection of the second  
scalar state will serve as a crucial test of the present model.

PACS/ keywords: 11.15.-q, 12.40.-y, 14.40.-n/ Relativistic bound
state description of $q\bar q$ states in the top-mass region. Scalar $0^+$
$q \bar q$ state identified with new particle found with a mass of 126 GeV. 
Higgs-field not needed. 
\end{abstract}

In the study of fundamental forces hadronic and weak
interactions give access to the smallest systems of nature with the
existence of different flavour systems~\cite{PDG}. The
observation of states in the top-mass region (with a mass significantly larger
than the bottonium-system) is of particular interest, since in addition to $t
\bar t$ states this is the mass region of the heavy bosons of the weak
interaction, but also of the Higgs-boson and supersymmetric particles, 
predicted in extensions of the Standard Model of particle physics~\cite{PDG}
(SM). Therefore, experimentally large efforts have been made to study this
mass region in detail. Recenty, a new particle has been
discovered~\cite{resnew} at LHC, which has been interpreted with large
confidence as the Higgs-boson. Evidence for supersymmetric particles has not
been found. 

The present experimental situation requires a critical view of the SM, a sum
of first order gauge field theories for the description of the 
electromagnetic, weak and strong interaction. Although this model describes
many particle properties, it is an effective theory with parameters and
assumptions, which have to be understood in a more fundamental theory.
Among the parameters of the SM is the electric
coupling constant $\alpha\sim 1/137$, the number of flavour families and the
masses of quarks and leptons. Big problems of this model are further the
understanding of massive neutrinos and the relation to gravitation, which
should be based also on particle properties. In a fundamental theory all these
features should be understood. Therefore, extensions of the SM to explain only
mass or flavour (as by the Higgs-mechanism and supersymmetry) have to be
viewed critically, if they are not part of a fundamental theory.   

From the general observation that nature is finite a
fundamental theory may be finite and contain higher order terms.  
However, there is the general belief that the only possible theories
to describe fundamental forces are first order gauge field theories. Divergent
higher order theories are not renormalisable, whereas 
other higher order theories have been found to lead to non-physical
results~\cite{highd}. But the latter theories cannot be a principal
problem, if a physical Lagrangian can be found, which respects all basic
features of a relativistic theory, as gauge invariance and energy-momentum
relation.

Recently, a finite theory based on a second order extension of the QED
Lagrangian by boson-boson coupling has been developed~\cite{Moneu1}, in which a
rather fundamental description of the electric force in light atomic systems
is achieved. In this formalism all parameters needed are constrained by
self-consistency conditions, so that a description without free parameters is
obtained. Even the magnitude of the coupling constant $\alpha_{QED}$ is
deduced (which is not understood in QED). 
Importantly, within this formalism not only the electric interaction between
hadrons and leptons can be described, but also the structure of the individual
particles, requiring the assumption of massless elementary fermions
(quantons). In this framework confinement, creation of bound states as well as
the existence of different flavour systems in hadrons~\cite{Moneu2} is
understood. No other theory is needed to understand the 
masses and the flavour degree of freedom.  
In this approach the quarks in the SM can be understood as effective fermions 
with masses given by eigenvalues in a boson-exchange potential. Likewise, the
heavy gauge bosons may be considered also as effective bosons, which cannot be 
detected experimentally.  

The used Lagrangian is of the form  
\begin{equation}
\label{eq:Lagra}
{\cal L}=\frac{1}{\tilde m^{2}} \bar \Psi\ i\gamma_{\mu}D^{\mu}
D_{\nu}D^{\nu}\Psi\ -\ \frac{1}{4} F_{\mu\nu}F^{\mu\nu}~,   
\end{equation}
where $\tilde m$ is the reduced mass and $\Psi$ a two-component
massless fermion (quanton, q) field $\Psi=(\Psi^+$ $\Psi^o$) and $\bar
\Psi= (\Psi^-$ $\bar \Psi^o$) with charged and neutral part. Vector boson
fields $A_\mu$ are contained in the covariant derivatives
$D_{\mu}=\partial_{\mu}-i{g} A_{\mu}$ and the Abelian field 
strength tensor $F^{\mu\nu}=\partial^{\mu}A^{\nu}-\partial^{\nu}A^{\mu}$. 
Generalised couplings to the charge ($g=g_c$) and spin ($g=g_s$) of quantons
have to be considered. A detailed discussion of the formalism is given in
ref.~\cite{Moneu1, Moneu2}.

Contributions to stationary solutions can be studied by evaluating fermion 
matrix elements $M_{ng}=\bar \psi(p')~V_{ng}(q)~\psi(p)$ with two potentials
$V_{2g}(q)$ and $V_{3g}(q)$, which are due to coupling of two ($2g$) and three
boson ($3g$) fields in the Lagrangian. The potential $V_{2g}(q)$ has been
identified with the confinement 
potential in hadrons, whereas $V_{3g}(q)$ can be considered as second order
boson-exchange potential. In r-space these potentials are given in the form  
\begin{equation}
V_{2g}(r)= \frac{\alpha^2\tilde m <r^2_{w_s}>F_{2g}}{4}\ \Big
(\frac{d^2 w_s(r)}{dr^2} + 
  \frac{2}{r}\frac{d w_s(r)}{dr}\Big )\frac{1}{\ w_s(r)}\ . 
\label{eq:vb}
\end{equation}
and
\begin{equation} 
\label{eq:vqq}
V^{s,v}_{3g}(r)= -\frac{\alpha^3 \hbar}{\tilde m} \int dr'
w_{s,v}^2(r')\ v_v(r-r')~.   
\end{equation}
These potentials involve bosonic (quasi) wave functions\footnote{leading to
  boson (quasi) densities $w^2_{s,v}(q)$ with 
  dimension $[GeV]^2$.} $w_{s}(r)$ and $w_{v}(r)$ of scalar and vector
character, respectively, whereas $v_v(r)$ can be regarded as boson-exchange
interaction $v_v(r)=-\hbar w_{v}(r)$. $F_{2g}$ is a Fourier transformation
factor due to the transformation of the boson kinetic energy to the potential
$V_{2g}(r)$.  
Both potentials~(\ref{eq:vb}) and (\ref{eq:vqq}) give rise to binding of
fermions, but the potential~(\ref{eq:vqq}) can be regarded also as bosonic
matrix element, with a binding of bosons by the interaction
$v_v(r)$. This yields a boson binding energy $E_b$.

The bosonic wave functions $w_{s}(r)$ and $w_{v}(r)$ give rise to two
states with quantum numbers $J^\pi =1^-$ and fermion wave functions
$\psi_{s,v} (r)\sim w_{s,v}(r)$, which are normalised to~1. 
The wave function $w_s(r)$ is used of the form
\begin{equation}
\label{eq:wf}
w_s(r)=w_{s_o}\ exp\{-(r/b)^{\kappa}\} \ , 
\end{equation} 
whereas $w_v(r)$ is written by 
\begin{equation}
\label{eq:spur}
w_{v}(\vec r)=w_{v_o}~(w_s(r)+\beta R\ dw_s(r)/dr) \ .  
\end{equation}
The factors $w_{(s,v)_o}$ are obtained from the normalisation $2\pi \int r
dr\ w_{s,v}^2(r)=1$. Further, $\beta R=-(\int r^2dr~w_s(r))/(\int
r^2dr~[dw_s(r)/dr])$ ensures orthogonality of the fermion wave
functions and cancellation of spurious motion $<r_{w_s,w_v}>=0$ for bosons. 

In addition there are two p-states (with $J^\pi=0^+$) with similar wave
functions.  Here, angular momentum-spin fractions
$(<\frac{1}{2} \frac{1}{2}|~L=1~ S_{gg}~|~0^+>/<\frac{1}{2}
\frac{1}{2}|~L=0,2~ S_{gg}~|~1^->)^2$ have to be taken into account, where
$S_{gg}$ is the spin coupling of the two bosons in $w_s(r)$ and $w_s(r)$. This
yields spin reduction factors for the binding energies of $0^+$ states, 
estimated to be $(2/3)^2$ and $(3/5)^2$ for $w_s(r)$ and $w_v(r)$, respectively. 

For a self-consistent determination of the parameters $\kappa$, $b$ and
$\alpha$ geometrical boundary conditions and energy-momentum relations are
needed. Geometric boundary conditions arise from the requirement that for the most
strongly bound $1^-$ state of the system the interaction should take place
inside the bound state volume. This leads to a similar form of the fermionic
and bosonic wave functions $\psi_{s,v}(r)\sim w_{s,v}(r)$ and 
\begin{equation}
\label{eq:conr}
{c}\ w^2_s(r) \sim |V^v_{3g}(r)| \ . 
\end{equation}

The mass of the system is defined by 
\begin{equation}
\label{eq:mass}
M_{n_{s,v}}=-E_{f_{s,v}}^{3g}+E_{f_n}^{2g} \ , 
\end{equation}
where $E_{f_{s,v}}^{3g}$ are negative binding energies in $V^{s,v}_{3g}(r)$ 
and $E_{f_n}^{2g}$ positive binding energies in $V_{2g}(r)$ (for
$n=1$ the index $n$ is dropped). For the binding in $V^{s}_{3g}(r)$ the
total energy of the system is not increased, the negative
fermion and boson binding energies $E_f^s$ and $E_g$ have to be compensated by
the root mean square momenta of the corresponding potentials, giving rise to
an energy-momentum relation   
\begin{equation}
<q^2_{V_{3g}}>^{1/2}+<q^2_{v_v}>^{1/2} = -(E_f^s+E_g) \ . 
\label{eq:massq}
\end{equation}
A last constraint arises from the confinement potential~(\ref{eq:vb}), see
ref.~\cite{Moneu2}, which gives 
\begin{equation}
Rat=\frac{\hbar^2}{\tilde m^2 <r^2_{w_s}>} =1 \ .
\label{eq:ravb}
\end{equation}

Altogether there are four constraints, orthogonality, relations
(\ref{eq:conr}), (\ref{eq:massq}) and (\ref{eq:ravb}), by which the
parameters $\kappa$, $b$ and $\alpha$ are unambiguously determined. The fact
that there are no other parameters in the entire formalism indicates clearly
that a fundamental theory is constructed. Nevertheless, there are small
ambiguities arising from the used forms of the wave functions in
eqs.~(\ref{eq:wf}) and (\ref{eq:spur}), which gives rise to estimated
uncertainties up to about 10 percent. 

Different flavour\footnote{the term flavour is kept from the quark
  model} systems are characterized in the present approach by a
different slope parameter $b$ only. Therefore, all flavour systems should have
a rather similar structure with two $1^-$ states, a very narrow low mass state
and a wider state at much larger mass, which is expected to decay rapidly to
two mesons, baryons or leptons. Also for the 
top system this is expected. Therefore, the observed $t\bar t$ peak at
about 350 GeV, which decays to two mesons or two ``single-top''
states~\cite{stop}, has to be identified with the high mass 1$^-$ state. The
low mass $1^-$ state should have a mass about a factor 4 smaller, where the
only state is $Z$(91.2 GeV). 

Here it should be recalled that $Z$(91.2 GeV) has been interpreted in the past
as gauge boson of the neutral weak interaction. However, as discussed above, 
particles (gauge bosons and quarks) needed in the effective theories of the SM
should be considered as effective particles, which may not be identified with
real physical 
states. This allows to interpret $Z$(91.2 GeV) as $q\bar q$ state. This 
is not inconsistent with the measured decays of this state into
hadrons and leptons, if the calculated width is in agreement with the sum of 
experimental decay widths (smaller than the total width of 2.5 GeV). 

--------

By applying the above formalism to $q\bar q$ states in the top-mass
region, a boson-density with a mean radius square of about $10^{-5}$ fm$^2$ is
required from a vacuum potential sum rule~\cite{Moneu1,Moneu4}. This yields
a fundamental $1^-$ state with a mass in the order of 80-100 GeV. 
By adjusting the parameters $b$, $\kappa$  and $\alpha$ by the constraints
discussed above, the potentials $V_{3g}(r)$ and $V_{2g}(r)$ are well
determined. Results on the radial dependence of densities and potentials
are given in fig.~1. In the upper part
the interaction $v_v(r)$ is given by the solid line. Compared to a Coulomb like
potential there are no divergencies for $r\to 0$ and $\infty$, consistent
with the requirement of a finite theory. 

\begin{table}
\caption{Results for the top-system in comparison with the
  data~\cite{PDG,resnew}. Masses are given in GeV, $b$ in fm, and mean 
  radius squares in fm$^2$. } 
\begin{center}
\begin{tabular}{lc|cc||cc}
~~~System & states & $M_s$ & $M_v$ & $M^{low}_{exp}$ &
  $M^{high}_{exp}$    \\   
\hline
vector (1$^-$) & $Z$, $t\bar t$ & 91.2 & 350 & 91.2$\pm$0.1 &
350$\pm$10  \\  
\hline
scalar (0$^+$) & new & 41 & 126 & ~ & 126$\pm$0.8   \\  
\end{tabular}

\begin{tabular}{ccc|c}
 $\kappa$ & $b$ &  $\alpha$ & $<r^2_{w_s}>$  \\ 
\hline
1.4  & 4.69 10$^{-3}$ & 2.61  &  1.63 10$^{-5}$ \\  
\end{tabular}
\end{center}
\end{table}
In the middle part a comparison of the density $w_s^2(r)$ (dot-dashed line)
with the potentials $V^s_{3g}(r)$ (dashed line) and $V^v_{3g}(r)$ (solid line)
is made. We see that condition~(\ref{eq:conr}) for the vector potential is
rather well fulfilled. The deduced parameters and radii are
given in table~1. As expected, the low mass $1^-$ state can be identified
with $Z$(91.2 GeV), whereas the mass of the second $1^-$ state was found to be
about 330 GeV, which is at least 20 GeV smaller than the $t\bar t$ state
observed experimentally~\cite{PDG}. 
This default can be cured easily by a small modification of the boson wave 
function $w_v(r)$. Replacing in eq.~(\ref{eq:spur}) the derivative
$dw_s(r)/dr$ by a form $dw_s(r)/dr +
c~d^2w_s(r)/dr^2$ with a tiny amplitude $c$ of 6 10$^{-4}$ $fm^2$, a value
of $M_v$ of about 350 GeV is obtained consistent with the experimental $t\bar
t$ peak. The root mean square momenta are found to be 
$<q^2_{V_{3g}}>^{1/2}$=109 GeV and $<q^2_{v}>^{1/2}$=231 GeV, yielding a sum
of 340 GeV. Further, $E_g$ was found to be -249 GeV leading to
$E^s_{f} + E_g$= -340 GeV. This shows that the energy-momentum
relation~(\ref{eq:massq}) is fulfilled.   

In fig.~2 the potential $V_{2g}(r)$ is given, which has the typical
form of the 'confinement' potential $V_{conf}=-\alpha/r+l\cdot r$ deduced from
potential models. However, in the present case this potential is very weak and
gives only a small contribution to the mass in the order of 0.02-0.04 GeV. For
lighter flavour systems (in particular for charmonium and bottonium) excited
states in the confinement potential have been found. Here, their masses
are only 0.25, 0.45 and 0.63 GeV above the low mass $1^-$ state. Therefore,
within the experimental width of 2.5 GeV these states cannot be observed.

In the lower part of fig.~2 the Fourier transform of the confinement
potential $T_{2g}(q)$ is shown, which is directly related to the 
mass distribution and width of the low mass $1^-$ state in question. 
The numerical Fourier expansion of this potential depends
strongly on the interpolation limits and detailed radial grid, and
can be well approximated by a Gaussian. This peak becomes extremely
narrow, if a high resolution in r and q together with integration to large
radii is used. With logarithmic interpolation and integration up to 0.25 fm a
width of less than 1.5 GeV is obtained, as shown in fig.~2, which is already
smaller than the observed width of $Z$(91.2 GeV) of 2.5 GeV. For integration
up to even larger radii a still narrower peak is observed, indicating that the
real width is extremely small.
  
Mass distributions due to the potential 
$V_{3g}(r)$ are given by the Fourier transform of the kinetic energy
distributions $T_{3g}(r)=\frac{1}{2} <r^2>({d^2 V_{3g}(r)}/{dr^2} +
\frac{2}{r}\ {dV_{3g}(r)}/ {dr})$. These give rise to very broad
distributions, which are shown by dot-dashed lines for the low mass $1^-$
state in the upper part and for the high mass state in the lower part of
fig.~3. This shows clearly that the confinement potential alone is responsible
for the observation of narrow $q\bar q$ states, but these small peaks are
found on top of a large 'background' contribution from the potential
$V_{3g}(r)$. This makes a detection of these states very difficult, 
as also found experimentally.

Concerning $0^+$ states, using the spin reduction factors given above one
state is predicted with a mass of about 41 GeV, 
the other with a mass of 126 GeV, which is in agreement with the mass of
the new particle.

The correctness of these results can be checked directly by realising that the
present formalism can be considered also as a fundamental theory of the electric
interaction in light atoms~\cite{Moneu1}. This has the consequence that many
features and characteristics of bound states should be relatively
similar in hydrogen and the top-system. So, the two $1^-$ states, $Z$(91.2
  GeV) and $t\bar t$(350 GeV) may be related to the 1s and 2s levels in H,
with a mass ratio $M_{Z(91.2~GeV)}/M_{t\bar t(350~GeV)}$ quite similar to the ratio
of binding energies $E_f(2s)/E_f(1s)$ in hydrogen. The new $0^+$ states in the
top-system should then be compared to the 2p and 3p states in H. In
particular, the mass ratios between $0^+$ and $1^-$ states
$M_{s,v}(0^+)/M_{s,v}(1^-)$ should be the same as the ratio of binding
energies between corresponding p and s states, since these quantities depend
only on angular momentum-spin coupling coefficients, see above. However, this
should be valid only for the binding energies in $V_{3g}(r)$. The relative
strength of $V_{2g}(r)$ is drastically different in both cases, with very small
binding energies $E^{2g}_f$ in the top system but about
10-40 \% of $E^{3g}_f$ for hydrogen. Since the relative strength of
$V_{2g}(r)$ to $V_{3g}(r)$ is affected by the Fourier transformation factor
$F_{2g}$, larger uncertainties are expected in $E^{3g}_f$
for hydrogen, whereas such errors are negligible for the top-system. 

Inspecting the ratio of binding energies for $0^+$ and $1^-$ states in the
analysis in ref.~\cite{Moneu1}, between the 2p and 1s levels in H a ratio
$E_f^{3g}(2p)/E_f^{3g}(1s)$ of 0.31 is found. By 
lowering $E_f^{3g}(1s)$ to -14.6 eV and increasing $E_f^{3g}(2p)$ to about -5
eV (which is within the estimated errors) this
ratio becomes 0.34. This is in reasonable agreement with the spin reduction
factor of 0.36 estimated for $M_{0^+(126~GeV)}/M_{t\bar t(350~GeV)}$. For the 3p and
2s levels in H a ratio $E_f^{3g}(3p)/E_f^{3g}(2s)$ of 0.44 is found,
which is the same as estimated for the ratio $M_{0^+(41~GeV)}/M_{Z(91.2~GeV)}$. 
This shows indeed a consistent picture of the two very different systems and
confirms the $0^+$ $q\bar q$ assignment of the new resonance at 126
GeV. However, to demonstrate the full applicability of the present formalism 
it will be important to find the second scalar state at about 41 GeV.
\vspace{0.7cm}

In summary, a fundamental (parameter free) description of the hadronic
interaction has been applied to the mass region of top-states. Two $1^-$
$q\bar q$ states have been predicted, which are in good agreement with states
observed 
experimentally. $Z$(91.2 GeV) has to be interpreted as the low mass $1^-$
$q\bar q$ top-state and not as a gauge boson. Its calculated width is very
small and consistent with the experimental widths. The high mass state is
identified with the $t\bar t$ peak at about 350 GeV, which decays
dominantly into two mesons or baryons.
Further, two $0^+$ $q\bar q$ states are found, one with  
a mass in agreement with the recently discovered scalar state at 126
GeV. This indicates that this state can be interpreted as scalar $q\bar q$
state and does not require an exotic interpretation as Higgs-boson.
A second scalar state is predicted with a lower mass of about 41 GeV, which
should be searched for in high energy experiments. Its detection
can be considered as a crucial test of the present model. 

As a general conclusion, quarks and massive gauge bosons required in
effective SM theories should be considered as effective
particles, which cannot be observed experimentally. Furthermore, particles
needed in extensions of the SM, as the Higgs-particle, should be viewed also
as effective particles. Thus, apart from photons real particles may exist only
in the form of hadrons and leptons or in the form of more complex systems. 
\vspace{0.5cm}

\newpage
\begin{figure} [ht]
\centering
\includegraphics [height=15cm,angle=0] {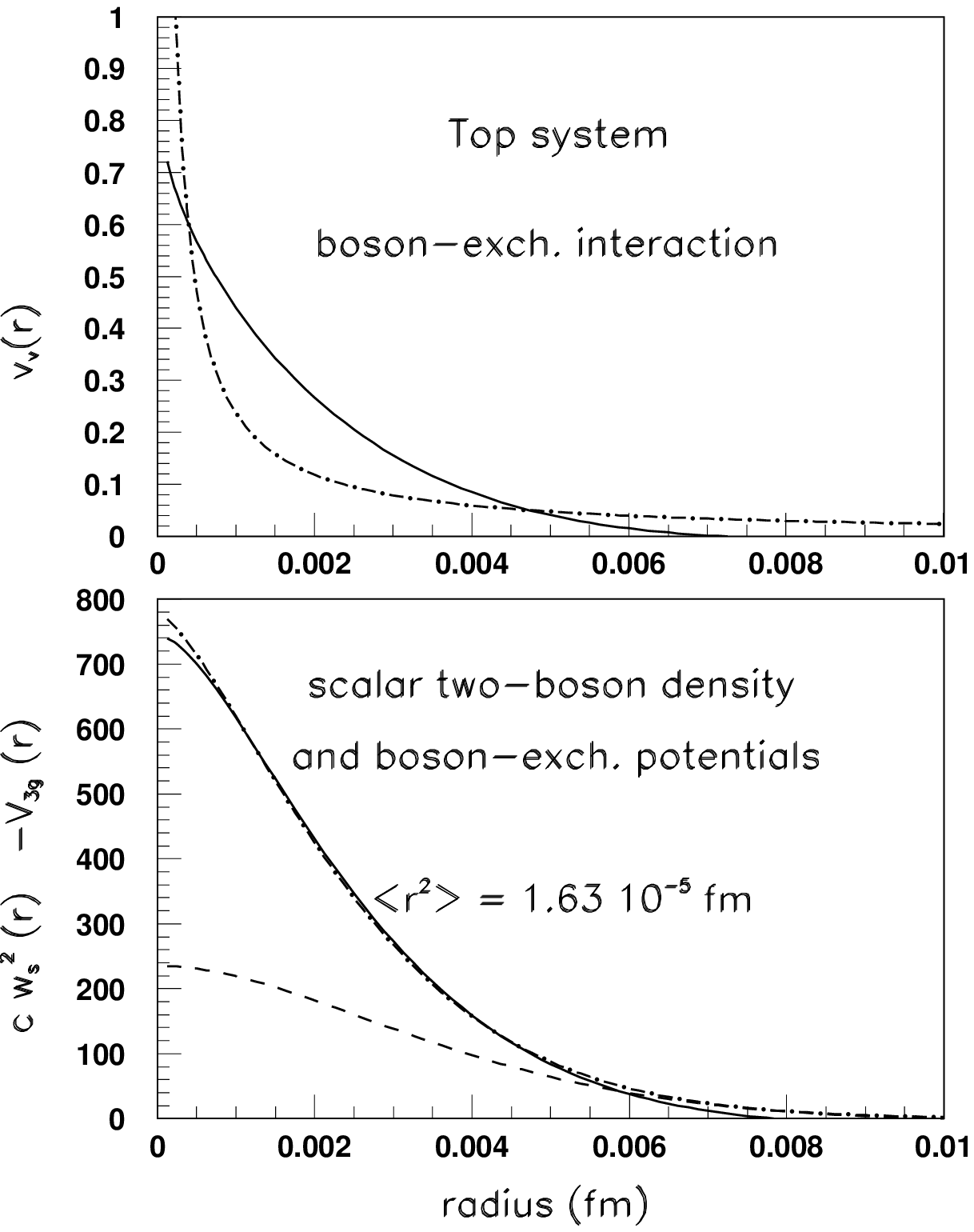}
\caption{Self-consistent solution for vector $q\bar q$ states in
  the top-mass region.
  \underline{Upper part:} Interaction $w_v(r)$ given by solid 
  line in comparison with a Coulomb like potential (dot-dashed line).
 \underline{Lower part:} Bosonic density $w_s^2(r)$ given by dot-dashed line,
 potential $|V^{v}_{3g}(r)|$ (solid line) and $|V^{s}_{3g}(r)|$ shown by
 dashed line. } 
\label{fig:g1ex2t}
\end{figure}

\begin{figure} [ht]
\centering
\includegraphics [height=15cm,angle=0] {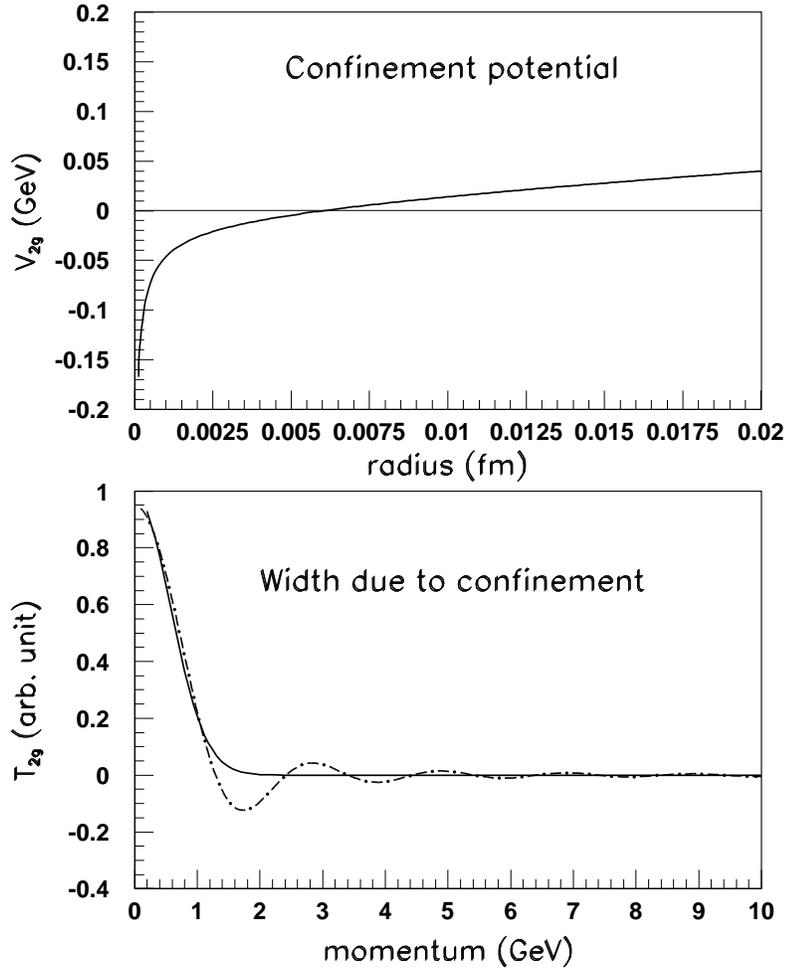}
\caption{Confinement potential $V_{2g}(r)$ (upper part) and Fourier transform
  (lower part) given by dot-dashed line. The solid line corresponds to a
  Gaussian form with a full width at half maximum of $\sim$1.5 GeV. Using 
  increasingly larger radial limits in the Fourier expansion, the width of 
  the peak reduces further.  }
\label{fig:confinetop}
\end{figure} 
\begin{figure} [ht]
\centering
\includegraphics [height=15cm,angle=0] {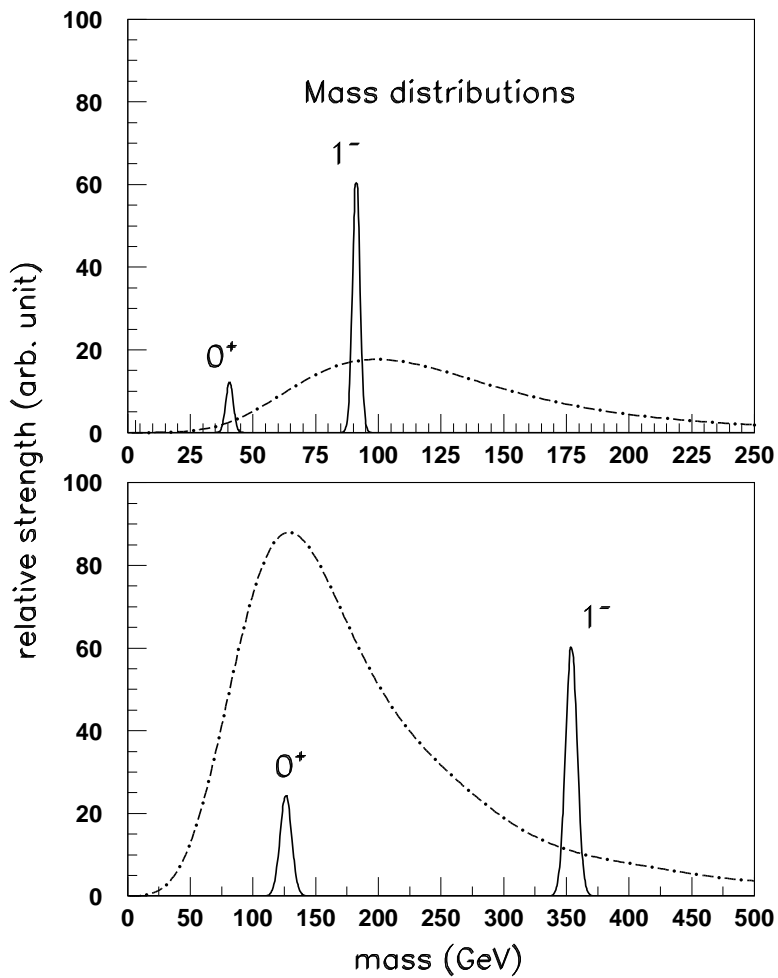}
\caption{Mass distributions for vector and scalar states.
  The pronounced peaks given by solid lines are due to the Fourier transforms
  of $V_{2g}(r)$ (with their widths arbitrarily enlarged), whereas the
  momentum distributions due to $V_{3g}(r)$ give 
  rise to the wide dot-dashed distributions (shown only for $1^-$ states).  
  The $0^+$ state with a mass of 120-130 GeV can be identified with the new
  scalar state found in Atlas and CMS data~\cite{resnew}. } 
\label{fig:sigma2t}
\end{figure} 

\end{document}